\documentclass[a4paper,11pt]{article}
\pdfoutput=1 

\usepackage{jcappub} 

\usepackage[T1]{fontenc} 

\title{\boldmath Bose--Einstein condensates in charged 
black--hole spacetimes}

\author[a]{El\'ias Castellanos,}
\author[b]{Juan Carlos Degollado,}
\author[c]{Claus L\"ammerzahl}
\author[d]{Alfredo Mac\'ias}
\author[c]{and Volker Perlick}

\affiliation[a]{Mesoamerican Centre for Theoretical Physics, Universidad Aut\'onoma de Chiapas, Carretera Zapata Km. 4, Real del Bosque (Ter\'an), 29040, Tuxtla Guti\'errez, Chiapas, M\'exico.}
\affiliation[b]{Instituto de Ciencias F\'isicas, Universidad Nacional Aut\'onoma de M\'exico,
Apdo. Postal 48-3, 62251, Cuernavaca, Morelos, M\'exico.}
\affiliation[c]{ZARM, Universit\"at Bremen, Am Fallturm, 28359 Bremen, Germany.}
\affiliation[d]{Departamento de F\'{\i}sica, Universidad Aut\'onoma Metropolitana--Iztapalapa, PO.Box 55-534, Mexico D.F. 09340, M\'exico.}
\emailAdd{ecastellanos@mctp.mx}
\emailAdd{jcdegollado@ciencias.unam.mx}
\emailAdd{laemmerzahl@zarm.uni-bremen.de}
\emailAdd{amac@xanum.uam.mx}
\emailAdd{perlick@zarm.uni-bremen.de}

\abstract{We analyze Bose--Einstein condensates 
on three types of spherically symmetric and static charged black-hole spacetimes:
The Reissner-Nordstr{\"o}m spacetime, Hoffmann's Born-Infeld black-hole spacetime, and
the regular Ay{\'o}n-Beato--Garc\'ia spacetime. The Bose-Einstein condensate is 
modeled in terms of a massive scalar field that satisfies a Klein-Gordon
equation with a self-interaction term. The scalar field is assumed to be uncharged 
and not self-gravitating. If the mass parameter of the scalar field is chosen
sufficiently small, there are quasi-bound states of the scalar field that may
be interpreted as dark matter clouds. We estimate the size and the total energy of 
such clouds around charged supermassive black holes and we investigate if their
observable features can be used for discriminating between the different types
of charged black holes.}

\begin{document}
\maketitle
\flushbottom

\section{Introduction}
\label{sec:intro}
\indent
Scalar fields appear in many situations in physics. A particularly important 
idea is to use scalar fields as candidates for dark matter \cite{DM,DM1,DM2,DM3,DM4}.
According to this idea dark matter consists of a certain type of
spin-zero bosons, known as Weakly Interacting Massive Particles (WIMPs), axions or other,
depending on the specific model, 
which have to be viewed as hypothetical because they have not been observed so far.
The bosonic character of
these particles, using the theory of relativistic Bose gases \cite{JB,LP}, 
also opens the door for the existence of scalar field dark matter in the 
form of Bose-Einstein condensates \cite{BoehmerHarko2007,ure,TonaRev}.

As we believe that most, if not all, galaxies harbor a supermassive black hole
at the center, the idea of modelling dark matter by scalar fields naturally
leads to the question of whether there are bound or quasi-bound states
of scalar fields near black holes. For the case of a Schwarzschild black hole
and a massive scalar field without self-interaction, it was found that such
quasi-bound states exist \cite{xxx,BarrancoEtAl2014}. More precisely, it was demonstrated 
that there are spherically symmetric field configurations, satisfying the boundary
conditions of no flux coming in from infinity or from the horizon, that persist
for a very long time. Exact solutions that satisfy these boundary conditions
cannot persist forever; they have to decay in the course of time. However, 
this is not a problem for modelling dark matter clouds around black holes 
in terms of  such scalar field configurations as long as the decay time is in the order of
the age of the Universe. Whereas in the two quoted articles the scalar field was
treated as a test field on the Schwarzschild background, numerical studies have
also been carried out for the case of a self-gravitating cloud \cite{QU,BarrancoEtAl2017}.

If one wants to pursue the idea of modelling dark matter as a Bose-Einstein
condensate, a self-interaction term has to be taken into account. In a fully relativistic
setting one has to add such a self-interaction  term to the massive Klein-Gordon equation
which results in an equation that is very similar to the (non-relativistic) Gross-Pitaevskii
equation. On Schwarzschild and Schwarzschild-de Sitter spacetimes, this equation and its
potential for modelling quasi-bound scalar clouds was discussed in Refs.\,\cite{NOS,NOS1}. 
In the present paper we want to extend this analysis to \emph{charged} black holes. More
precisely, we want to consider three types of charged black holes that arise from
different theories of electrodynamics, and we want to investigate if one can
discriminate between these black holes from the observation of quasi-bound clouds of
Bose-Einstein condensates. The three types of charged black holes are the Reissner-Nordstr{\"o}m
black hole, Hoffmann's Born-Infeld black hole \cite{Hoffmann1935} 
and the Ay{\'o}n-Beato--Garc{\'\i}a black hole \cite{ABG}.
All of them are spherically symmetric and static charged black holes. They arise from 
coupling to Einstein's field equation the standard Maxwell theory, the Born-Infeld 
theory \cite{BornInfeld1934} and another particular non-linear electrodynamical theory 
of the Pleba{\'n}ski class \cite{plebas70}, respectively.

While these three types of black holes have the same asymptotics far away from
the center, they have quite different features inside the horizon. The 
Reissner-Nordstr{\"o}m metric has a curvature singularity and a diverging
electric field strength at the center, while Hoffmann's black hole has a curvature 
singularity and a finite electric field strength there. The Ay{\'o}n-Beato--Garc{\'\i}a 
metric describes a \emph{regular} black hole, i.e., the metric has no curvature
singularity. The existence of regular black-hole solutions makes charged black
holes coupled to nonlinear electrodynamics particularly interesting from a 
conceptual point of view. The first regular black hole metric was brought 
forward by Bardeen \cite{Bardeen1968} in 1968 who, however, did not investigate 
the question of whether this metric was a solution to Einstein's field equation 
with a reasonable matter source. The properties of regular black holes were
further studied in 
Refs.\,\cite{Dymni92,BarrabesFrolov1996,Borde1997,Marsetal1996,bron2}.
The Ay{\'o}n-Beato--Garc{\'\i}a metric 
was the first regular black hole that was found as a solution to Einstein's field
equation with a nonlinear electrodynamical field as the source. Soon thereafter more such
solutions were found \cite{ABG1,Bronnikov2001,Burinskii02}; in particular it was shonw
that the Bardeen black hole is also such a solution, with a magnetic monopole at 
the center \cite{AyonGarcia2000}.

The fact that different electrodynamical theories lead to charged black
hole solutions with qualitatively quite different interior raises the 
question of whether these differences are observable from the outside.
Spacetimes of charged black holes may be probed with test particles and 
with light rays as has been demonstrated in several articles: 
The motion of (charged or uncharged) test particles in
the Reissner-Nordstr{\"o}m spacetime has been studied in great detail,
see 
\cite{Jaffe1922,Gackstatter1983,TeliPalaskar1984,Chaliasos2001,Hackmannetal08hd,GrunauKagramanova11}. 
For Hoffmann's Born-Infeld black hole, the geodesics have been 
investigated in Ref.\,\cite{Breton2002} and the light rays in 
Ref.\,\cite{Eiroa2006}.  
For geodesics in the Ay{\'o}n-Beato--Garc{\'\i}a metric we refer to 
Refs.\,\cite{KalamFarhadHossein2014,StuchlikSchee2015}. 

In addition to using particles or light rays, one may also use scalar fields
for probing different spacetime geometries. It is the purpose of this paper
to investigate if the quasi-bound states of scalar fields which have been
suggested for modeling dark matter clouds can be used for discriminating 
between different types of charged black holes.  

The outline of the paper is as follows: in Section \ref{sec:cbh} we briefly
review the basic properties of the charged black hole spacetimes under
consideration. In Section \ref{sec:gp} we consider, on these spacetimes, 
an uncharged scalar field that satisfies   the Klein-Gordon equation with 
a self-interaction term, we derive the corresponding Gross--Pitaevskii
equation and we analyze some properties of the effective potential .
In Secton \ref{sec:qb} we numerically construct quasi-bound states 
and we compare them for our three charged black-hole spacetimes.   
In Section \ref{sec:tf} we discuss if the Thomas-Fermi approximation is 
viable for this kind of quasi-bound states. In Section \ref{sec:conclusions}, 
we present the conclusions and some outlook.

\section{Spherically symmetric and static charged black holes}
\label{sec:cbh}

We consider spherically symmetric and static spacetimes where the metric, in 
standard sphe\-ri\-cal coordinates, is  given as
\begin{equation}\label{eq:g}
g_{\mu \nu} dx^{\mu} dx^{\nu} 
= -f(r) c^2 dt^{2}+\frac{1}{f(r)}dr^{2}+r^{2}(d \theta^{2}+\sin^{2}\theta
d\phi^{2}) \, .
\end{equation}
We compare three different types of such spacetimes all of which describe
charged black holes, but in different theories of electrodynamics. 
The first is the Reissner-Nordstr{\"o}m (RN) spacetime with
the metric function $f(r)$ equal to
\begin{equation}\label{eq:RN}
f_{\mathrm{RN}}(r)   = 1- \dfrac{2m}{r} + \dfrac{q^2}{r^2} \, .
\end{equation}
This is the unique spherically symmetric and static solution of 
Einstein's field equation coupled to standard Maxwell electrodynamics. 
The second one is Hoffmann's 
Born-Infeld (HBI) black-hole spacetime \cite{Hoffmann1935},
\begin{equation}\label{eq:HBI}
f_{\mathrm{HBI}}(r)   = 
1-\frac{2m}{r}
+ \dfrac{2}{\sigma ^2 r} \int _r ^{\infty}
 \Big( \sqrt{s^4+ \sigma ^2 q^2} -s^2 \Big) ds
\end{equation}
which is a solution of Einstein's field equation coupled to 
Born-Infeld electrodynamics \cite{BornInfeld1934}.
The third one is the Ay{\'o}n-Beato--Garc{\'\i}a (ABG) spacetime \cite{ABG},
\begin{equation}\label{eq:ABG}
f_{\mathrm{ABG}}(r)   = 
1-\frac{2mr^{2}}{(q^{2}+r^{2})^{3/2}}+\frac{q^{2}r^{2}}{(q^{2} +r^{ 2})^ {2}}
\end{equation}
which is a regular black-hole solution of Einstein's field equation
coupled to a certain non-linear electrodynamical theory of the Pleba\'nski 
class \cite{plebas70}. 
In all three cases, $m$ is the mass parameter and $q$ is the charge parameter, 
both of which have the dimension of a length. They are related to the ADM mass
$M$ and the electric charge $Q$ in SI units by
\begin{equation}\label{eq:MQ}
m = \dfrac{GM}{c^2} \, , \quad 
q= \dfrac{\sqrt{G} \, Q}{\sqrt{4 \pi \varepsilon _0} \, c^2}
\end{equation}
where $G$ is Newton's gravitational constant and $\varepsilon _0$ is the 
permeability of the vacuum.  The HBI metric involves a third
parameter, $\sigma$, which also has the dimension of a length. The Born-Infeld
theory postulates the existence of a constant of Nature with the dimension of
a magnetic field strength, $b$, and the parameter $\sigma$ is equal to 
\begin{equation}\label{eq:sigma}
\sigma = \dfrac{cM}{bQ} \, .
\end{equation}
For $b \to \infty$ the Born-Infeld theory approaches the standard Maxwell theory;
correspondingly, the HBI metric approaches the RN metric for $\sigma \to 0$.
For $\sigma \to \infty$ the HBI metric approaches the Schwarzschild metric. The
fact that to date the standard Maxwell theory is in agreement with all experiments
demonstrates that $b$ must be big. If we assume that $cb$ (which has the 
dimension of an electric field) is much bigger than the electric field of
our black hole in the entire domain $m \lessapprox r < \infty$, we have to
require that 
\begin{equation}\label{eq:sigmaest}
\sigma \ll m^3/q^2 \, .  
\end{equation}
The ABG metric is regular at the origin while the RN and the HBI solutions 
have a curvature singularity there. All three metrics describe black holes,
i.e., they have one or more horizons, if the parameters $m$, $q$ and $\sigma$ 
are chosen appropriately. Horizons are indicated by zeros of the metric function
$f(r)$. The RN metric has an inner and an outer horizon 
at radii $0 < r_{\mathrm{Hi}} < r_{\mathrm{Ho}} < \infty$ if $0 < |q| < m$. 
For $|q| =m$ the two horizons merge and for $|q| >m$ the singularity at the
origin is naked. Similarly, the ABG metric features two horizons if
\begin{equation}\label{eq:qc}
0 < |q| < q_c \, , \quad q_c \approx 0.634 \, m \, .
\end{equation}
For $|q| > q_c$ the ABG metric is
a regular metric without a horizon. As we want to compare different black-hole
spacetimes with the same parameters $m$ and $q$ we will restrict to values
of $q$ and $m$ that satisfy the inequality (\ref{eq:qc}). It is then easy to check
that the HBI metric describes a black hole (with one or two horizons) for all
values of $\sigma$.  

While the RN, the ABG and the HBI metrics are quite different near the 
origin, they have the same asymptotics far away from the center, 
\begin{equation}\label{eq:asympt} 
f(r) = 1 - \dfrac{2m}{r}+\dfrac{q^2}{r^2} + O (r^{-3}) \, .
\end{equation}
In the following we consider the region between the (outer) horizon and infinity,
where $f(r) >0$, and it is our goal to investigate if scalar field condensates
in this region show significant differences for the three cases.

\section{The Gross--Pitaevskii--like equation}
\label{sec:gp}

The Klein--Gordon equation for a complex test--scalar field $\Phi$ with a
scalar potential $V(\Phi)$ in a spacetime with metric $g_{\mu \nu}$ can be written as follows:
\begin{equation}
\label{eq:KG0}
\frac{1}{\sqrt{-g}} \partial_{\mu} 
\Bigl(\sqrt{-g} g^{\mu \nu} \partial_{\nu}\Phi \Bigr)
-\frac{d\,V(\Phi \Phi ^*)}{d\,\Phi^{*}}=0,
\end{equation}
where $g$ is the determinant of the metric and an upper  star means complex conjugation. 
We consider a potential of the form
\begin{equation}
\label{eq:MEX}
V(\Phi \Phi^*)=\mu^{2}\Phi\Phi^{*}+\frac{\lambda}{2}(\Phi\Phi^{*})^{2}
\end{equation}
which allows to interpret the scalar field as a Bose-Einstein condensate of
some bosonic particles. Here $\mu$ is, as usual, 
the scalar field mass parameter which equals the inverse Compton wavelength
of the particles, i.e.
\begin{equation}\label{eq:Compton}
\mu = \dfrac{M_{\Phi} c}{\hbar} 
\end{equation}
where $M_{\Phi}$ is the mass of the particles, and $\lambda$ is the 
self--interaction coupling constant which equals, up to a numerical factor,  
the scattering length $a_s$ of the particles, $\lambda = 16 \pi a_s$.
Note that $\lambda$ has the dimension of a length while $\mu$ has the
dimension of an inverse length. Here we assume that the scalar field 
describes uncharged particles. For particles with a charge $Q_{\Phi}$
one would have to couple in an electromagnetic field by the usual minimal 
replacement rule, $\partial _{\mu} \Phi \mapsto \partial _{\mu} \Phi + i \hbar ^{-1}
Q_{\Phi} A_{\mu} \Phi$ where $A_{\mu}$ is the electromagnetic potential. 

We are interested in (approximate) solutions of the Klein-Gordon equation
on a spacetime of the form (\ref{eq:g}) that are spherically symmetric and 
harmonic in time,
\begin{equation}\label{eq:omega}
\Phi (t,r) =
e^{i\,\omega\,t}\,\frac{u(r)}{r}
\end{equation}
with a \emph{real} frequency $\omega$. In this case, after some algebraic 
manipulations the Klein--Gordon equation reduces to a Gross--Pitaevskii--like 
equation,
\begin{equation}
\Bigg(-  \dfrac{d^2}{dr_*^2}  + V_{\rm eff} (r) +
\lambda_{\rm eff} (r) \,\dfrac{|u(r)|^2}{r^2} \Bigg)\,u (r)
= \dfrac{\omega^{2}}{c^2} \,u (r) \, .
\label{eq:KGGP}
\end{equation}
Here $r_*$ denotes the \emph{tortoise coordinate} which is defined by
the equation
\begin{equation}
d\,r_*=\frac{d\,r}{f(r)} \, .
 \label{eq:r_es} 
\end{equation}
The tortoise coordinate runs over the entire real line when $r$ runs from the 
(outer) horizon to infinity. Moreover, we have introducced
%
%
the effective potential
\begin{equation}
V_{\rm eff}(r)= f(r)\,\,\left(\mu^2 + \frac{f'(r)}{r} \right)
\label{eq:Veff}
\end{equation}
and the effective self-interaction 
parameter
\begin{equation}
\lambda_{\rm eff} (r) =\lambda\,f(r) \, .
\label{eq:lameff}
\end{equation}
The fact that $\lambda_{\rm eff}$ depends on $r$ via the metric function
$f(r)$ reflects the influence of the spacetime geometry on the self-interaction, 
cf. Refs.\,\cite{NOS,NOS1}. By comparison with the standard Gross-Pitaevskii 
equation we see that $\omega^{2}/c^2$ may be identified with an effective chemical 
potential. Of course, in the relativistic case $\omega$ occurs quadratic, rather 
than linear as in the standard Gross--Pitaevskii equation, because the Klein-Gordon
equation involves a second time derivative.

Introducing the tortoise coordinate $r_*$, viewing the radial coordinate $r$
as an implicitly given function of $r_*$, is convenient for analyzing the differential
equation. However, when comparing
different black-hole spacetimes we will always work with the coordinate $r$, rather than 
with $r_*$. The reason is that the radius coordinate $r$ always has the same geometric
meaning of giving the area $A(r)$ of the sphere at radius $r$ by the usual formula 
$A(r) = 4 \pi r^2$. By contrast, the tortoise coordinate depends on the chosen spacetime 
and admits no general interpretation in terms of a measurable quantity.

With every solution $\Phi$ of the Klein-Gordon equation (\ref{eq:KG0})  we associate the
current
\begin{equation}\label{eq:j}
j^{\mu} = i \, \alpha \, g^{\mu \nu} \big( \Phi ^* \partial _{\nu} \Phi - \Phi \partial _{\nu} \Phi ^* \big)
\end{equation}
where $\alpha$ is a real constant.
It follows immediately from (\ref{eq:KG0}) that this current is conserved,
\begin{equation}\label{eq:consj}
\dfrac{1}{\sqrt{-g}} \partial _{\mu} \Big( \sqrt{-g} \, j^{\mu} \Big) = 0 \, .
\end{equation}
With $\alpha$ chosen appropriately, (\ref{eq:consj}) is to be interpreted as the conservation 
law of the particle number. 
If $\Phi$ is of the form of (\ref{eq:omega}), with a real frequency $\omega$, we may
choose 
\begin{equation}\label{eq:alpha}
\alpha = \dfrac{c^2}{2 \omega} \, .
\end{equation} 
Then the only non-zero components of the current are
\begin{equation}\label{eq:jt}
j^t =  \dfrac{| u (r) |^2}{ f(r) r^2} \, ,
\end{equation}
\begin{equation}\label{eq:jr}
j^r =  \dfrac{i c^2 f(r)}{2 \omega r^2} \Big( u(r)^* \dfrac{du(r)}{dr}- u(r) \dfrac{du(r)^*}{dr} \Big) 
\end{equation}
and the conservation law (\ref{eq:consj}) reduces to
\begin{equation}\label{eq:consj2}
\partial _t j^t + \dfrac{1}{r^2} \partial _r \Big( r^2 j^r \Big) = 0 
\end{equation}
where the two terms on the left-hand side are separately equal to zero. If we multiply (\ref{eq:consj})
with $r^2 \mathrm{sin} \, \vartheta \, dr \, d \vartheta \, d \varphi$ and integrate over a spherical 
shell with inner radius $r_1$ and outer radius $r_2$ we get
\begin{equation}\label{eq:intconsj}
\dfrac{d}{dt} N_{r_1r_2}  + J_{r_2}-J_{r_1} = 0 
\end{equation}
where
\begin{equation}\label{eq:Nr}
N_{r_1r_2} = 4 \, \pi \int _{r_1} ^{r_2} \dfrac{|u(r)|^2}{f(r)} \, dr 
\end{equation}
is the number of particles in the shell and 
\begin{equation}\label{eq:J}
J_{r} = \dfrac{2i \pi c^2}{\omega} \Big( u(r)^*  f(r) \dfrac{du(r)}{dr}-
u(r)  f(r) \dfrac{du(r)^*}{dr} \Big)
\end{equation}
is the number flux through the sphere of radius $r$.

In addition to the conservation law for the particle number, there is also a 
conservation law of energy associated with our scalar fields. This follows from 
the fact that, as our spacetime is static, the Klein-Gordon equation (\ref{eq:KG0}) 
can be equivalently rewritten as 
\begin{equation}\label{eq:Econ}
\dfrac{1}{\sqrt{-g}} \partial _{\mu} \Big( \sqrt{-g} T^{\mu}{}_t \Big) = 0
\end{equation}
with the energy-momentum tensor 
\begin{equation}
T_{\rho \sigma} = \dfrac{\hbar c}{\mu} \Big( \partial _{\rho} \Phi \partial _{\sigma} \Phi ^* 
+ \partial _{\rho} \Phi ^* \partial _{\sigma} \Phi  - g_{\rho \sigma} \big(
g^{\lambda \nu} \partial _{\lambda} \Phi \partial _{\nu} \Phi ^* + V( \Phi ^* \Phi) \big) \Big)
\, .
\end{equation}
For fields of the form of (\ref{eq:omega}) this conservation law simplifies to
\begin{equation}\label{eq:Econ2}
\partial _t \big( -  T^t{}_t \big) + \dfrac{1}{r^2} \partial _r \big(-  r^2 T^r{}_t \big) = 0
\end{equation}
with  both terms on the left-hand side separately equal to zero, 
where 
\begin{equation}\label{eq:Ttt}
- T^t{}_t = \dfrac{\hbar \, c}{\mu} \Bigg( 
\left( \dfrac{\omega ^2}{c^2 f(r)}
+ \mu ^2  + \dfrac{\lambda \, | u(r) |^2}{2 \, r^2}
\right) \dfrac{|u(r)|^2}{r^2}
+ f(r) \left| \dfrac{d}{dr}\left(  \dfrac{u(r)}{r} \right) \right| ^2  \Bigg)
\end{equation}
is the energy density and
\begin{equation}\label{eq:Trt}
- T^r{}_t = \dfrac{i \hbar c \omega}{\mu \, r^2} \Big( u(r)^* f(r) \dfrac{du(r)}{dr}
-u(r)  f(r) \dfrac{du(r)^*}{dr} \Big) 
\end{equation}
is the energy flux. Note that the energy flux is zero if $u(r)$ is real.  In analogy to
(\ref{eq:Nr}), the total energy in a shell between radii $r_1$ and $r_2$ is
\begin{equation}\label{eq:Er}
E_{r_1r_2} = - 4 \pi  \int _{r_1} ^{r_2} T^t{}_t \, r^2 \, dr
\end{equation}
\[
=
\dfrac{ 4 \, \pi \, \hbar \, c}{\mu} \int _{r_1} ^{r_2} 
\Bigg( \left(  \dfrac{\omega ^2}{c^2 f(r)}
+ \mu ^2 + \dfrac{\lambda \, | u(r) |^2}{2 \, r^2} \right)
 \, | u(r) |^2
+  r^2 f(r)  \left| \dfrac{d}{dr} \left(  \dfrac{u(r)}{r} \right) \right| ^2  \Bigg) \, dr \, .
\]

\begin{figure}[!htb]
\centering
\includegraphics[width=0.8\textwidth,origin=c,angle=0]{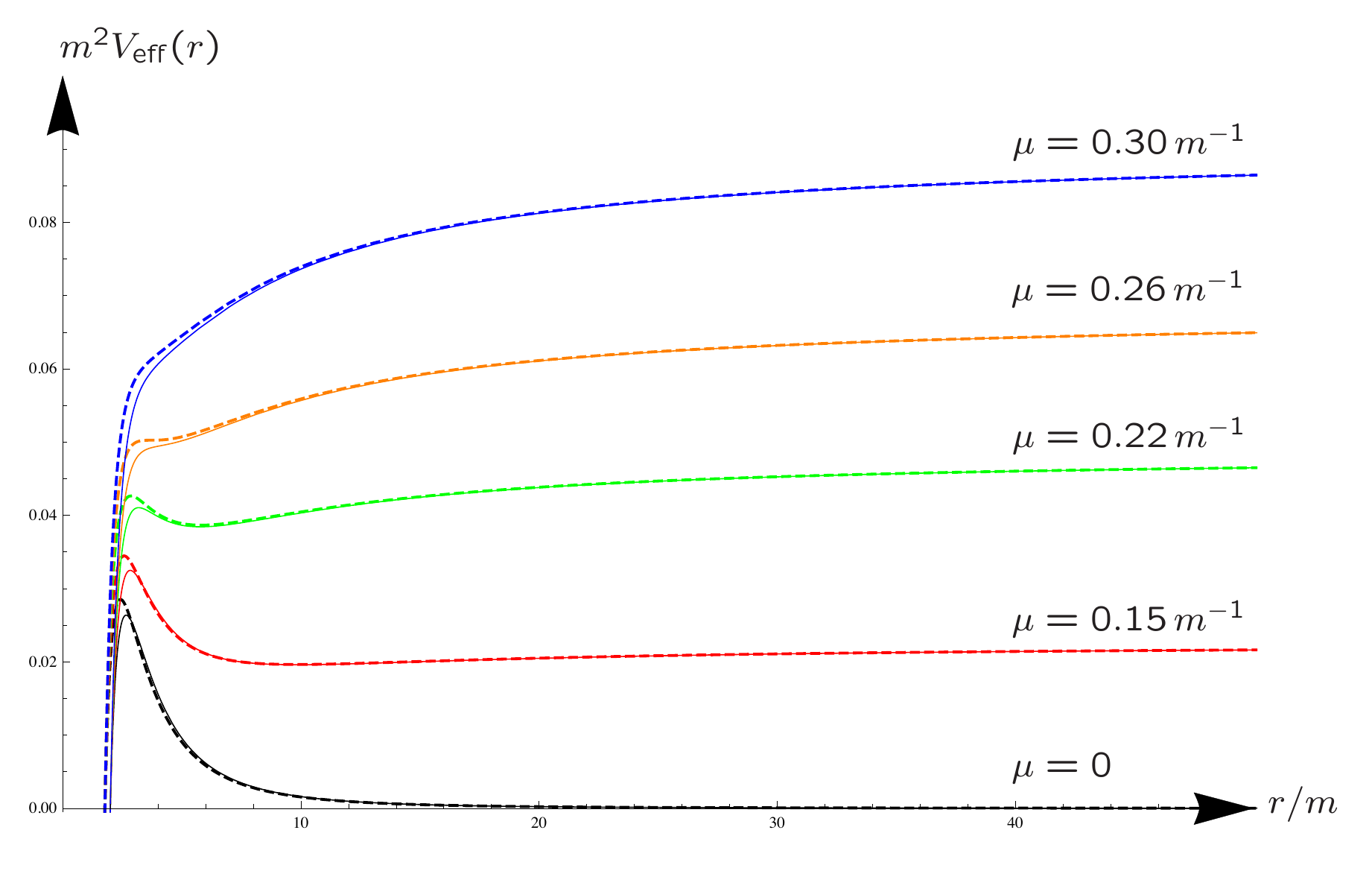}
\caption{\label{fig:Veff1} The effective potential $V_{\rm eff} (r)$ for the 
HBI black hole for different values of the scalar field mass parameter $\mu$. 
The charge is chosen as $q=0.634 \, m$ because we want to compare 
with an extremal ABG black hole. We give $r$ in units of $m$ and $V_{\mathrm{eff}}$
in units of $m^{-2}$. For each value of $\mu$ we have plotted
the limiting cases $\sigma =0$ (dashed) and $\sigma = \infty$ (solid) which
correspond to the RN black hole and the Schwarzschild black hole, respectively.
For any other value of $\sigma$ we get a curve that lies between the dashed and
the solid one. For realistic values of $\sigma$ that satisfy (\ref{eq:sigmaest}), 
the HBI case is practically undistinguishable from the RN case.
}
\end{figure}

\begin{figure}[!hbt]
\centering
\includegraphics[width=0.8\textwidth,origin=c,angle=0]{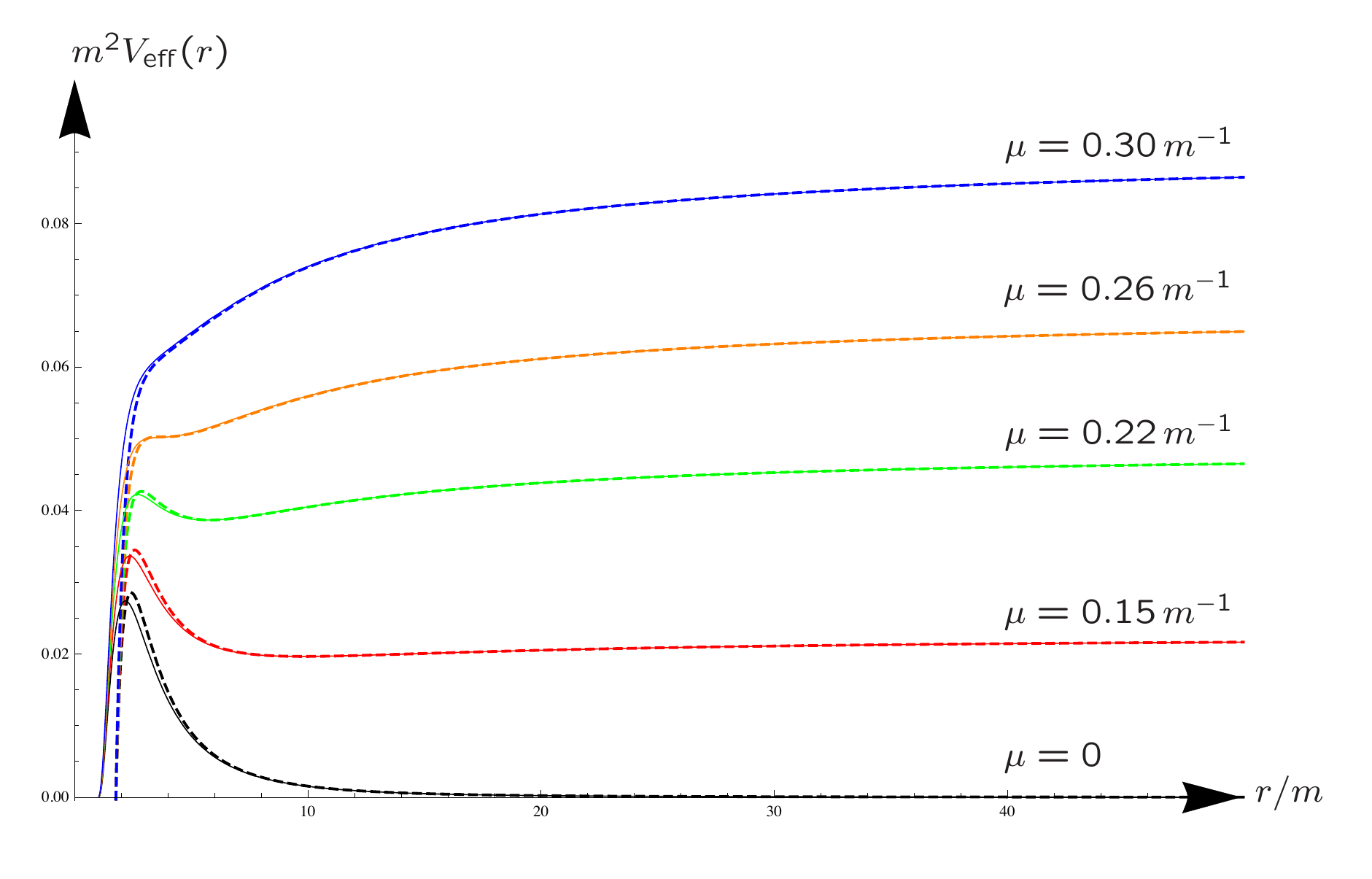}
\caption{\label{fig:Veff2} The effective potential $V_{\rm eff} (r)$ for the 
ABG black hole (solid) and the RN black hole (dashed) for different values
of the scalar field mass parameter $\mu$. The charge of the black hole is 
$q=0.634 \, m$, as in Figure \ref{fig:Veff1}. One sees a difference between 
the ABG and the RN case, in particular as to the position of the horizon; 
however, even for this highly charged black hole the minima and maxima of the
ABG potential are very close to those of the RN potential.  
}
\end{figure}

When experimenting with Bose-Einstein condensates in the laboratory  one
chooses an oscillator potential, or some similar potential that admits
a minimum and increases to infinity in all spatial directions, for 
trapping the condensate. Our potential $V_{\mathrm{eff}} (r)$ has a different
shape. However, for an appropriate choice of the mass parameter $\mu$ 
it features a local minimum which can provide some partial trapping. 
We speak of a \emph{partial} trapping because the potential has a maximum of 
finite height; so the scalar field is not perfectly trapped near the
local minimum but it may tunnel through the potential barrier and,
actually, decay in the course of time. Therefore, as long as we assume
that there is no flux of the scalar field coming in from infinity or from
the horizon, we cannot expect the existence of solutions that are 
strictly of the form of (\ref{eq:omega}) with a real $\omega$. However,
\emph{approximate} solutions of this kind may exist. We will refer to them
as to {\emph{quasi-bound}} states.

For investigating the existence of quasi-bound states it will be of crucial 
importance to determine for which values of $\mu$ the potential $V_{\mathrm{eff}}$ 
admits a minimum. We have plotted $V_{\mathrm{eff}}$ in Fig.\,\ref{fig:Veff1}
for the HBI metric and in Fig.\,\ref{fig:Veff2} for the ABG metric, both in 
comparison to the RN metric, for different values of the scalar field mass 
parameter $\mu$. The potential goes to zero at the horizon and it approaches 
the value $\mu ^2$ for $r \to \infty$. There is a critical value $\mu _c$ of
the parameter $\mu$ such that for $0 < \mu < \mu _c$ the potential has
a maximum and a minimum while for bigger values of $\mu$ it is monotonically
increasing. As the existence of a minimum is necessary for the existence of 
quasi-bound states, for $\mu =0$ and for $\mu > \mu _c$ such states do not 
exist. The value of $\mu _c$ depends on $q$ and it is different for RN, HBI
and ABG black holes, but it is always near $0.25 \, m^{-1}$. For a 
supermassive black hole, $m > 10^6 \mathrm{km}$, 
this value of $\mu$ corresponds to a particle mass $M_{\Phi}$ of not more than
$10^{-14} \mathrm{eV}/c^2$. So we need very light bosonic particles for 
producing condensates that may be (partially) trapped by our potential.

\begin{figure}[!htb]
\centering
\includegraphics[width=0.75\textwidth,origin=c,angle=0]{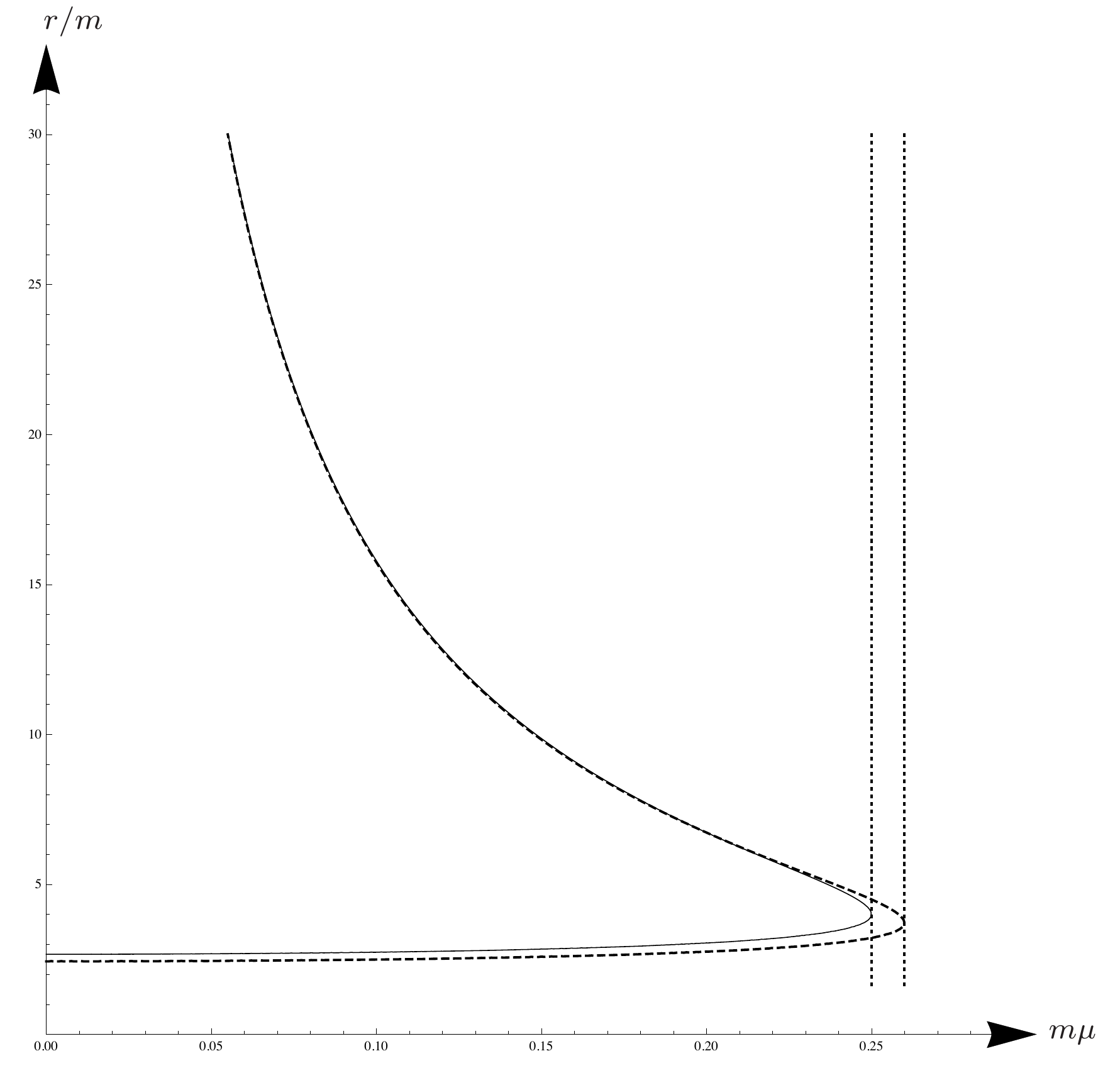}
\caption{\label{fig:Minmax1}
Plot of the equation $\frac{d}{dr}V_{\rm eff}=0$ for the HBI metric with
$q=0.634 \, m$. We have plotted the limiting cases $\sigma \to 0$ (dashed)
and $\sigma \to \infty$ (solid) which correspond to the RN metric and to
the Schwarzschild metric, respectively. For any other values of 
$\sigma$ the curve lies between these two ones.
For realistic values of $\sigma$ satisfying (\ref{eq:sigmaest}), the curve for 
the HBI black hole is practically undistinguishable from the curve for the
RN black hole.
}
\end{figure}

\begin{figure}[!htb]
\centering
\includegraphics[width=0.75\textwidth,origin=c,angle=0]{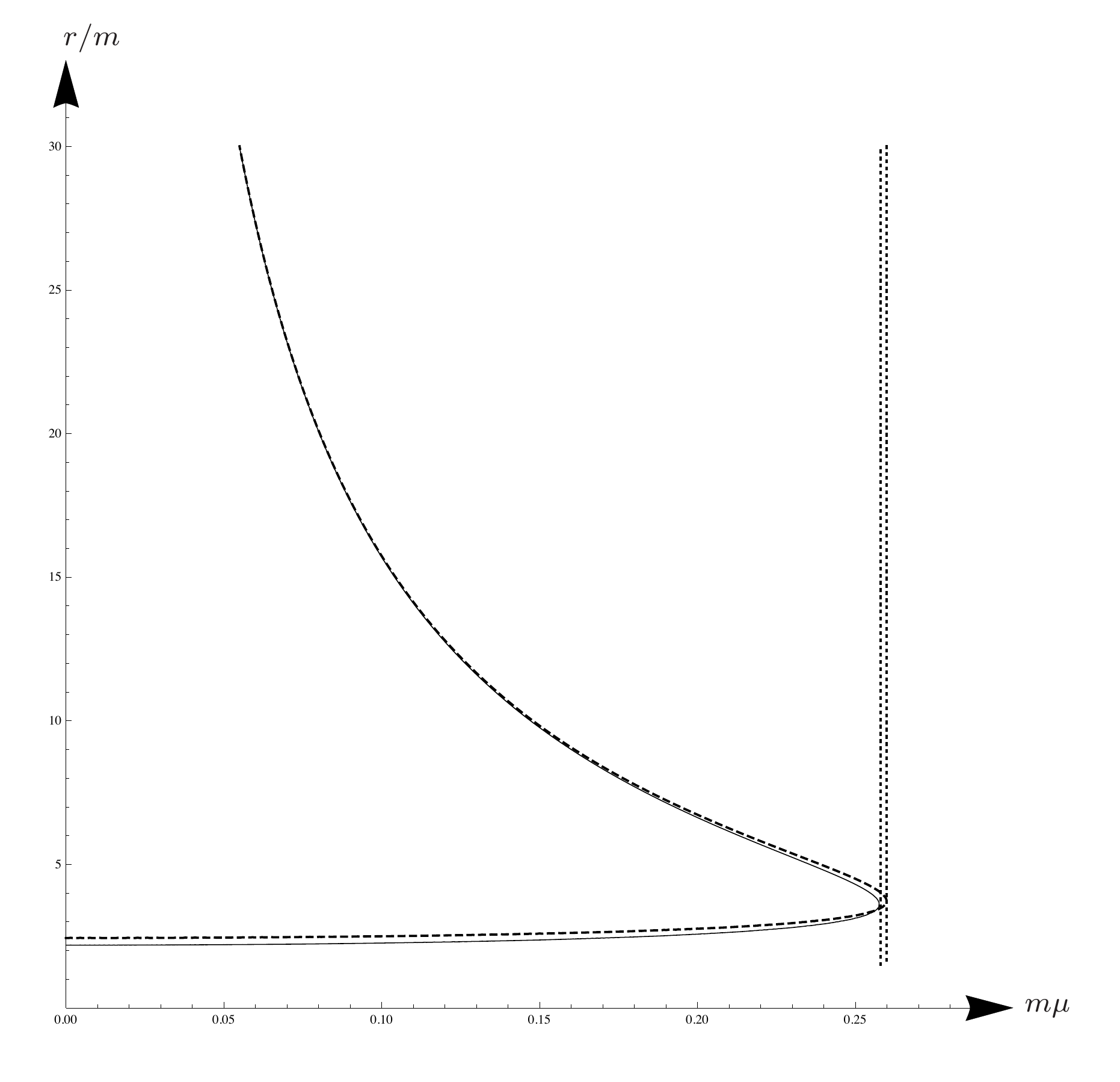}
\caption{\label{fig:Minmax2}
Plot of the equation $\frac{d}{dr}V_{\rm eff}=0$ for the ABG metric (solid)
and for the RN metric (dashed). As in Fig.~\protect\ref{fig:Minmax1}, the charge
parameter is chosen as $q = 0.634 \, m$. 
}
\end{figure}

For determining the  maxima and minima of $V_{\mathrm{eff}}$ we need to analyze the 
roots of the equation
\begin{equation}\label{eq:der_pot}
 \frac{d}{dr}V_{\rm eff} (r) =0\ .
\end{equation}
In the RN scenario, this yields a fifth order 
equation for $r$,
\begin{equation}
 6 q^4 - 15 m q^2 r + 8 m^2 r^2 + 4 q^2 r^2 - 3 m r^3 - \mu^2 q^2 r^4 +
  m \mu^2 r^5 = 0 \, , \,\,\,\,\,\,\,\,\,\,\,
\label{eq:der_potRN} \, .
\end{equation}
For the ABG metric it gives a 14th order equation and for the HBI metric it
gives an equation in terms of elliptic functions. 
In Figs.\,\ref{fig:Minmax1} and \ref{fig:Minmax2} we display the equation 
$\frac{d}{dr}V_{\rm eff}=0$ in the $\mu$-$r$ plane between the horizon and infinity,
where we give $\mu$ in units
of $m^{-1}$ and $r$ in units of $m$. If a vertical line, corresponding to a 
particular value of $\mu$, intersects this curve twice, there are real 
solutions, $r_{max}$ and $r_{min}$, for equation~(\ref{eq:der_pot})
outside of the horizon, indicating that the function $V_{\mathrm{eff}}$ 
has a maximum and a minimum. We see from the figures that this is true for 
$0 < \mu < \mu _c$, where the value of $\mu = \mu _c$ is indicated by a 
dotted vertical line in the figure. At this value the maximum and the minimum 
merge, forming a saddle. For $\mu > \mu _c$ the potential 
has no extrema. 

In this section we have seen that the effective potential in the ABG and in the HBI
case is very similar to that of the RN metric. Correspondingly, 
we expect that these different types of charged black holes admit very similar
scalar field configurations. In the next section we will see that this is, indeed,
true.

\section{Scalar quasi-bound states}
\label{sec:qb}

We have already emphasized that for a partial trapping $\mu$ must be chosen such
that the potential $V_{\mathrm{eff}}$ admits a local minimum, $0 < \mu < \mu _c$. 
For constructing quasi-bound states we then have to choose $\omega$ such that 
\begin{equation}\label{eq:mucon}
V_{\mathrm{eff}} ( r_{\mathrm{min}} ) < \frac{\omega^2}{c^2} < 
\mathrm{min} \big(\mu ^2 , V_{\mathrm{eff}} (r_{\mathrm{max}} )\big) \, .
\end{equation}
For the following it will be convenient to rewrite the Gross-Pitaevskii-like eqation
(\ref{eq:KGGP}) in the form
\begin{equation} 
\Bigg(-  \dfrac{d^2}{dr_*^2}  + V_{\rm eff} (r) +
f(r) \,\dfrac{|v(r)|^2}{r^2} \Bigg)\, v (r)
= \dfrac{\omega^{2}}{c^2} \, v (r) 
\label{eq:KGGP2}
\end{equation}
where
\begin{equation}\label{eq:v}
v(r) = \sqrt{\lambda} \, u(r) \, .
\end{equation}
Note that in terms of the function $v(r)$ the particle number (\ref{eq:Nr}) and the 
flux (\ref{eq:J}) are given by the equations
\begin{equation}\label{eq:Nr2}
\lambda \, N_{r_1r_2} = 4 \, \pi \int _{r_1} ^{r_2} \dfrac{|v(r)|^2}{f(r)} \, dr =
 4 \, \pi \int _{r_{*1}} ^{r_{*2}} | v (r)|^2\, dr_*
\end{equation}
and 
\begin{equation}\label{eq:J2}
\lambda \, J_{r} = \dfrac{2i \pi c^2}{\omega} \Big( v(r)^*  f(r) \dfrac{dv(r)}{dr}-
v(r)  f(r) \dfrac{dv(r)^*}{dr} \Big) =
\dfrac{2i \pi c^2}{\omega} \Big( v(r)^*   \dfrac{dv(r)}{dr_*}
- v(r)  \dfrac{dv(r)^*}{dr_*} \Big) \, ,
\end{equation}
respectively. In analogy to (\ref{eq:Nr2}), the expression  (\ref{eq:Er}) for the 
energy becomes
\begin{equation}\label{eq:Er2}
\lambda \, E_{r_1r_2} = 
\dfrac{ 4 \, \pi \, \hbar \, c}{\mu} \int _{r_{*1}} ^{r_{*2}} 
\Bigg( \left(  \dfrac{\omega ^2}{c^2 }
+ f(r) \mu ^2 + \dfrac{f(r) \, |v(r) |^2}{2 \, r^2} \right)
 \, |v(r) |^2
+  r^2  \left| \dfrac{d}{dr_*} \left(  \dfrac{v(r)}{r} \right) \right| ^2  \Bigg) \, dr_* \, .
\end{equation}

With $\mu$ and $\omega$ chosen appropriately, the partial trapping is reflected by 
the asymptotic behavior of solutions to (\ref{eq:KGGP2}): For big  $r$,
this equation can be approximated by
\begin{equation}\label{eq:GPasy1}
\dfrac{d^2 v(r)}{dr_*^2} \approx \Big( \mu ^2 - \dfrac{\omega ^2}{c^2} \Big) v(r)
\, , \quad r_* \to \infty
\end{equation}
while near the horizon we have
\begin{equation}\label{eq:GPasy2}
\dfrac{d^2 v(r)}{dr_*^2} \approx - \, \dfrac{\omega ^2}{c^2} \, v(r)
\, , \quad r_* \to - \infty \, .
\end{equation}
Correspondingly, there are solutions 
which exponentially decay for big valus of $r$,
\begin{equation}\label{eq:GPsol1}
v(r) \approx \gamma \, e^{- \sqrt{\mu ^2 -\omega ^2/c^2}\,  r_*} 
\, , \quad r_* \to  \infty 
\end{equation}
with a real constant $\gamma$
while real solutions have an oscillatory behavior near the horizon,
\begin{equation}\label{eq:GPsol2}
v(r) \approx \beta \, e^{-i \omega  r_* /c} 
+ \beta ^* \, e^{i \omega  r_* /c} 
\, , \quad r_* \to  - \infty 
\end{equation}
with a complex constant $\beta$. (\ref{eq:GPsol2}) is a superposition of an ingoing
and an outgoing particle flux. The ingoing one can be interpreted as a current of
particles that have tunnelled through the potential barrier and are falling towards
the black hole. The outgoing one is a hypothetical counter-current that is necessary
for providing a stationary solution. As no particles can come out of a
black hole, this counter-current cannot be expected to exist in Nature, so the 
solution that will actually be realised in Nature is not stationary but rather 
decaying in the course of time because of the particle flow that goes towards
the horizon. However, if the amplitude $|\beta |$ is small the solution may be 
considered as stationary over a long period of time. It is this kind of approximately
stationary, i.e. quasi-bound, solutions that we want to discuss in this section for 
the various types of black holes under consideration.

\begin{figure}[!hbt]
\centering
\includegraphics[width=0.7\textwidth,origin=c,angle=0]{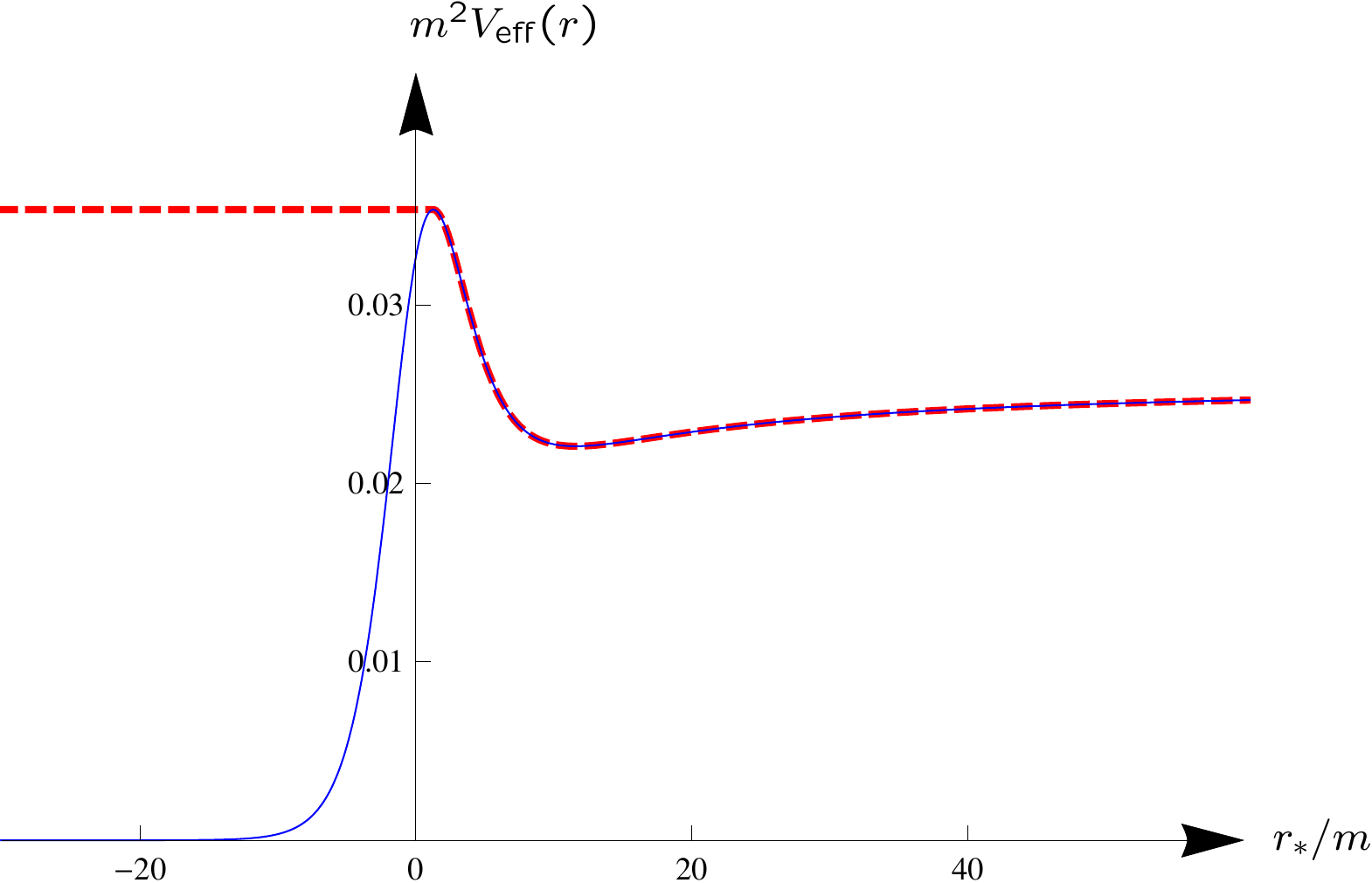}
\caption{\label{fig:VtV} The effective potential $V_{\rm eff} (r)$ (solid)
and the cut-off potential $\tilde{V}{}_{\mathrm{eff}}(r)$ (dashed) plotted against 
the tortoise coordinate $r_*$ for a RN black hole with $q = 0.634 \, m$
and $\mu = 0.16 \, m^{-1}$.}
\end{figure}

As  a formal means for switching off the tunneling we replace the
potential $V_{\mathrm{eff}}$ by a modified potential
\begin{equation}\label{eq:Weff}
\tilde{V}{}_{\mathrm{eff}} (r) = \, \left\{ \begin{matrix}
V_{\mathrm{eff}} (r_{\mathrm{max}} ) \: \: \qquad \text{if} \: \;  r\le r_{\mathrm{max}}
\\[0.4cm]
V_{\mathrm{eff}}(r) \: \: \quad \text{if} \: \:  r_{\mathrm{max}} < r < \infty
\end{matrix} \right.
\end{equation}
where $r_{\mathrm{max}}$ is the $r$ value where the potential takes its local maximum, see
Fig.~\ref{fig:VtV}.
In the potential $\tilde{V}{}_{\mathrm{eff}}$ we have stationary solutions in the strict sense,
i.e., solutions  $\tilde{V} (r)$ that fall off towards infinity and towards the horizon and are, thus, 
square-integrable. These solutions coincide on the interval $r_{\mathrm{max}} < r <
\infty$ with solutions $v(r)$ in the potential $V_{\mathrm{eff}}$; in the region between the
horizon and $r_{\mathrm{max}}$, however, the solutions $v(r)$ need the above-mentioned
unphysical counter-current for being stationary. We may view a stationary solution
$\tilde{v}(r)$ as a good approximation for a solution to our physical problem if the
particle number flux (\ref{eq:J2}) of this counter-current is sufficiently small. From the 
asymptotic formula (\ref{eq:GPsol2}) we read that this flux is given by 
\begin{equation}\label{eq:Jhor}
\lambda \, J_{r_* \to - \infty}  = 4 \pi c | \beta |^2 \, .
\end{equation}    
 
We will now construct such a quasi-bound state for a Reissner-Nordstr{\"o}m black 
hole with $q = 0.634 \, m$ and compare it to the other types of charged black holes 
afterwards. We choose $\mu = 0.16 \, m^{-1}$, 
see again Fig.~\ref{fig:VtV}, and $\omega /c = 0.15997 \, m^{-1}$. 
Choosing $\mu$ bigger  (i.e., closer to $\mu _c \approx 0.25 \, m^{-1}$)  gives a
lower potential barrier, i.e., it allows for more tunnelling; choosing $\omega /c$
bigger (i.e., closer to $\mu$) gives a condensate with a higher particle number.

\begin{figure}[!hbt]
\centering
\includegraphics[width=0.48\textwidth,origin=c,angle=0]{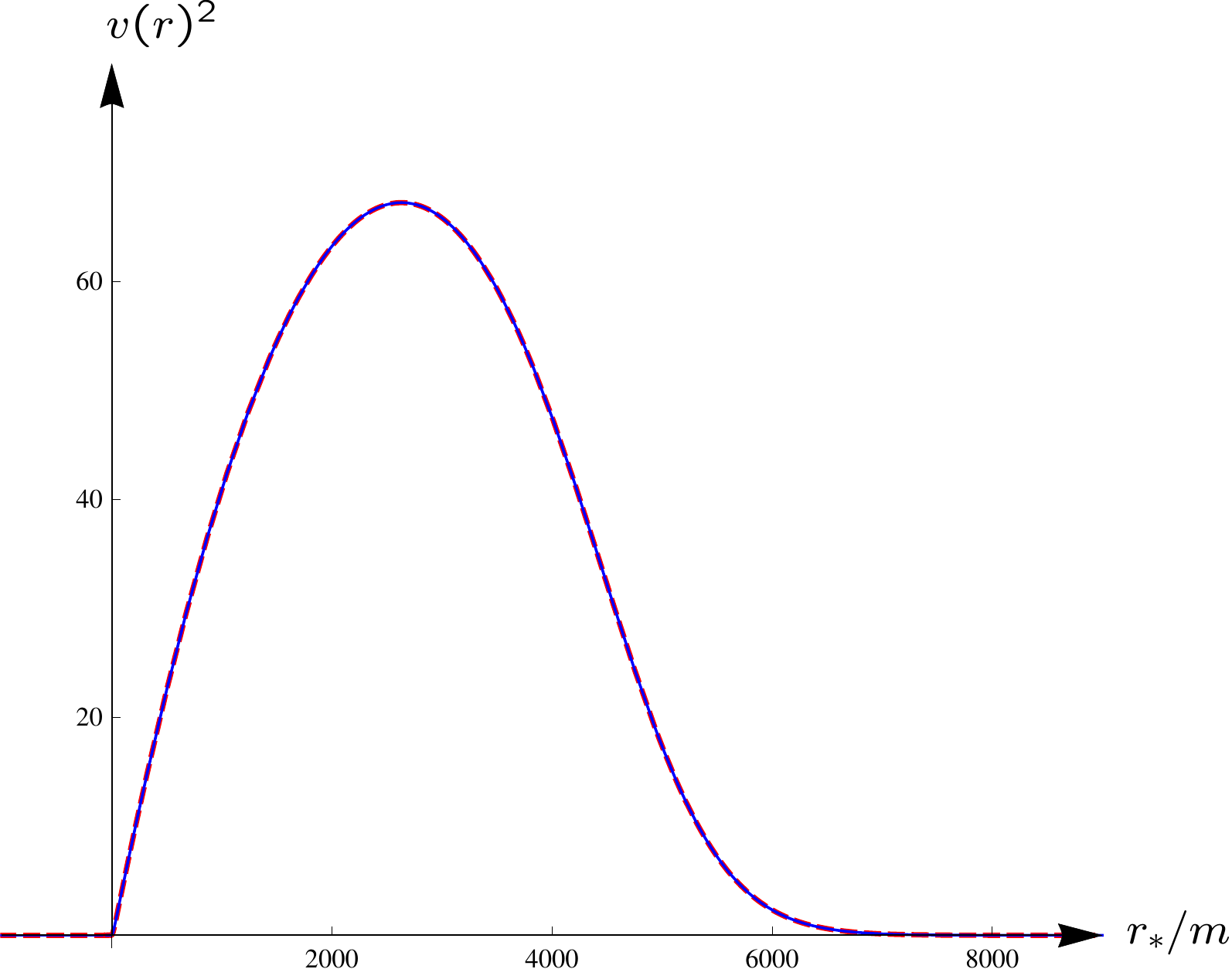}
\hspace{0.2cm}
\includegraphics[width=0.45\textwidth,origin=c,angle=0]{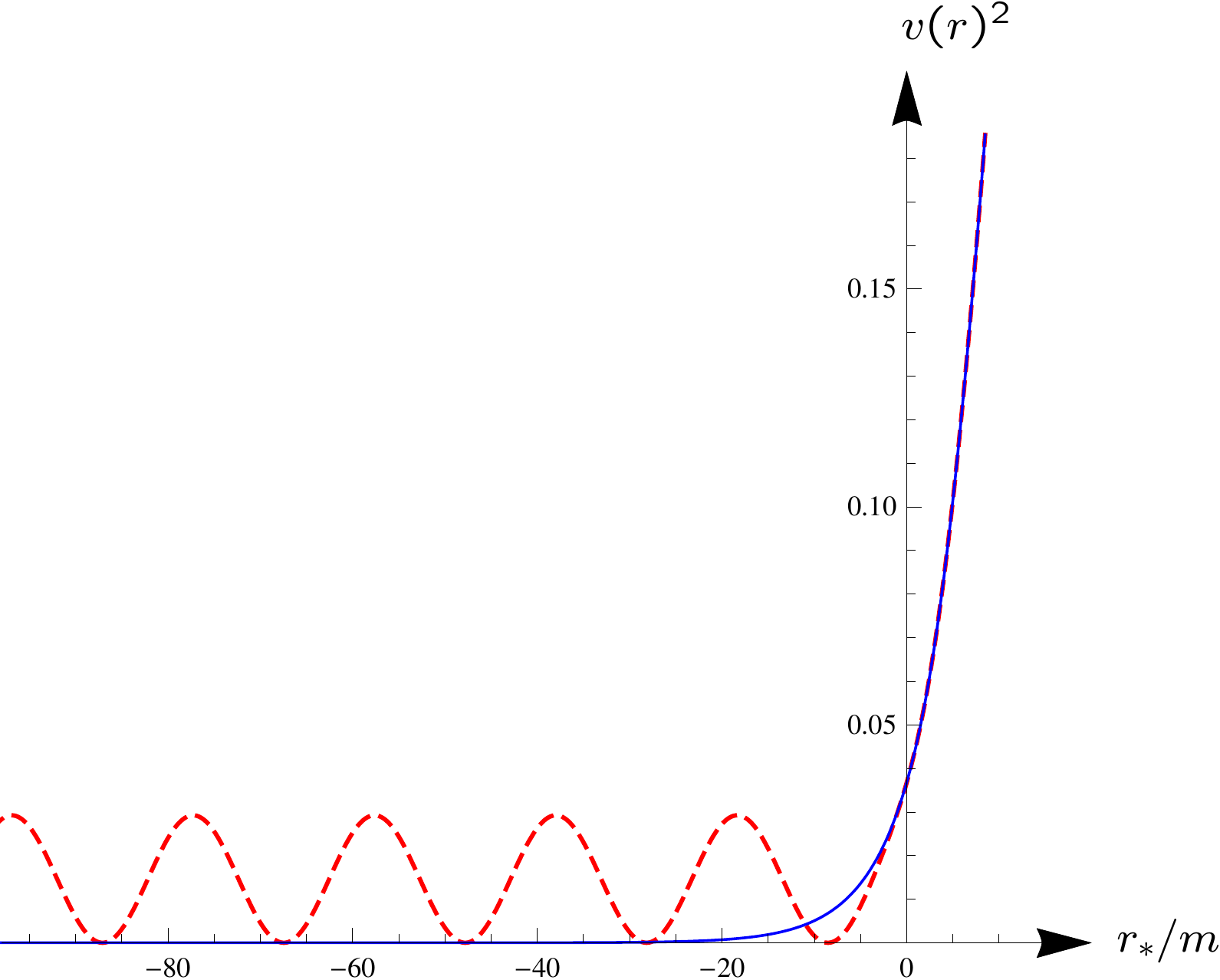}
\caption{\label{fig:vtv} Scalar field distributions $v (r)^2$
(dashed) and $\tilde{v}{}(r)^2$ (solid) for a RN black hole with 
$q = 0.634 \, m$, $\mu = 0.16 \, m^{-1}$ and 
$\omega/c = 0.15997 \, m^{-1}$, plotted against the tortoise coordinate.
$v(r)$ is a solution of (\protect\ref{eq:KGGP2}) with the potential $V _{\mathrm{eff}} (r)$ while 
$\tilde{v} (r)$ is a solution of the same equation with the cut-off potential
$\tilde{V} _{\mathrm{eff}} (r)$. The two solutions coincide on the interval 
$r_{\mathrm{max}} < r < \infty$. Towards the horizon, $\tilde{v} (r)$ falls off
exponentially whereas $v(r)$ oscillates according to (\protect\ref{eq:GPsol2}),
see the enlarged part of the plot on the right.
}
\end{figure}

We consider (\ref{eq:KGGP2}) with the cut-off potential $\tilde{V}{}_{\mathrm{eff}}$
instead of $V_{\mathrm{eff}}$. With $\mu$ and $\omega$ given, there is a unique
real and positive solution $\tilde{v} (r)$ of this equation that exponentially falls off towards  infinity and 
towards the horizon and has no zeros. We have determined this solution numerically, see Fig.~\ref{fig:vtv}.
As this solution is square-integrable, it yields a finite particle number $N$. By inserting
the numerical solution into (\ref{eq:Nr2}) we find
\begin{equation}\label{eq:N3}
\lambda \, N \approx 3 \times 10^6 \, m 
\end{equation}
where $m = GM/c^2$ is the mass parameter of the black hole.
On the interval $r_{\mathrm{max}} < r < \infty$ the function $\tilde{v} (r)$ coincides with 
a real solution $v(r)$ of (\ref{eq:KGGP2}), with the original potential $V_{\mathrm{eff}}$.
If extended beyond $r_{\mathrm{max}}$, this solution $v(r)$ approaches the horizon in
an oscillatory fashion according to (\ref{eq:GPsol2}), see again Fig.~\ref{fig:vtv}, so 
$v(r)$ is \emph{not} square-integrable. From the asymptotic behavior of the numerical 
solution we can read the value of $\beta$, according to (\ref{eq:GPsol2}); inserting $\beta$
into (\ref{eq:Jhor}) gives us the particle number flux $J_{r_* \to - \infty}$ that is necessary for 
compensating the loss by particles that tunnel towards the horizon,
\begin{equation}\label{eq:Jhor2}
\lambda \, J_{r_* \to - \infty} \approx 0.09 \, c \, .
\end{equation}    
The quotient
\begin{equation}\label{eq:T}
T = \dfrac{N}{J_{r_* \to - \infty}} = \dfrac{ \lambda N}{\lambda J_{r_* \to - \infty}} \approx   
 2.6 \times 10^6 \, \dfrac{m}{c}
\end{equation}
is a measure for the lifetime of the cloud. Note that $T$ is independent of $\lambda$, as long as
$\lambda >0$. (Our solutions do not have a finite limit for $\lambda \to 0$ because, by (\ref{eq:v}),
$u(r)$ goes to infinity in this limit.) Even for a supermassive black hole with $m \approx 10^{10} \, \mathrm{km}$,
such as the one at the center of M87,    (\ref{eq:T}) gives a lifetime of only $T \approx 35 \, 000$ years.
By astrophysical standards, this is not a very long lifetime. Bigger clouds with longer lifetime may
be constructed by choosing $\omega /c$ even closer to $\mu$. In analogy to (\ref{eq:N3}) we 
find the total energy $E$ of the cloud by inserting the numerical solution into (\ref{eq:Er2}), 
\begin{equation}\label{eq:Etot}
\lambda \, E \approx 2.2 \, \hbar \, c \, .
\end{equation}

For specifying numerical values for the particle number and for the energy it is 
necessary to specify $\lambda$. As an example, we choose $\lambda \approx 10^{-35} m$ 
which, for the above-mentioned  supermassive black hole with $m =GM/c^2 \approx 10^{10} \, \mathrm{km}$,
is equivalent to $\lambda \approx 10^{-25} \, \mathrm{km}$. This is not an unrealistic scattering
length for light dark matter candidates, cf. \cite{HarkoMocanu2012}. With this choice of $\lambda$,
(\ref{eq:N3}) yields $N \approx 10^{35}$ and (\ref{eq:Etot}) yields $E \approx 10^{-60} M c^2$.
So we see that the energy content of the cloud is tiny in comparison with the energy of the black hole. On the one hand, 
this confirms that the test-field approximation which we have used throughout is justified. On the
other hand, it demonstrates that the gravitational effect of a cloud with the chosen parameters would 
be practically unobservable. Nonetheless, the chosen parameters are appropriate for our main purpose:
We will demonstrate that, with these parameters, the resulting clouds in the BHI and ABG
spacetimes are virtually undistinguishable from the RN case, and that the differences are even
smaller for a more realistic choice of the parameters.

\begin{figure}[!hbt]
\centering
\includegraphics[width=0.75\textwidth,origin=c,angle=0]{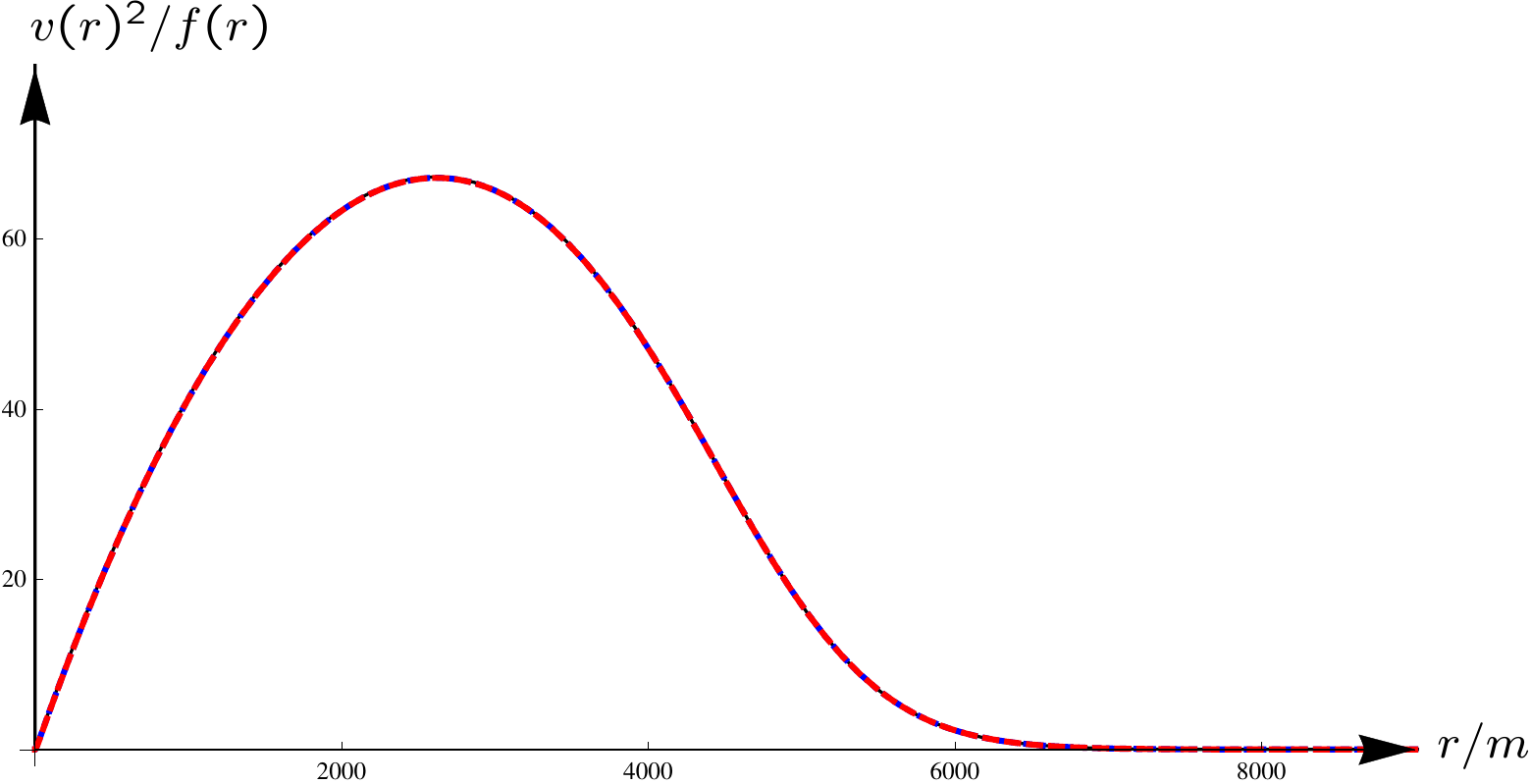}
\caption{\label{fig:3wigs} Density function $v(r)^2/f(r)$ for solutions of (\protect\ref{eq:KGGP2})
with $q=0.634m$, $\mu = 0.16 m^{-1}$ and $\omega /c = 0.15997 m^{-1}$, for the RN
black hole (solid) which is the limit of the HBI black hole for $\sigma \to 0$, for the limit of 
the HBI black hole for $\sigma \to \infty$ (dashed) and for the ABG black hole (dotted). 
The three graphs are lying on top of each other.   
}
\end{figure}

With the chosen values of $q = 0.634 \,  m$, $\mu = 0.16 \, m^{-1}$ and $\omega /c = 0.15997 \, m^{-1}$
we repeat the calculation that we have carried through for the RN black hole now for the HBI and for 
the ABG black holes. As the tortoise coordinate has a different geometric meaning in different
spacetimes, we plot the density distribution against the area radius function $r$ rather than against $r_*$,
see Fig.~\ref{fig:3wigs}. With respect to $r$, the number density is given up to a factor of $\lambda$
by the function $v(r)^2 / f(r))$, recall (\ref{eq:Nr}), so it is this function that we plot.
We see that the solutions in the three different black hole spacetimes are virtually undistinguishable.

We have found this result for the case that $q = 0.634  \, m$ which is the highest value of the charge
for which a comparison is possible. It is widely believed that the black holes that exist in Nature have
a considerably lower charge. Then the differences between the three types of black holes are even
smaller. Also, we have seen that, although we have chosen  $\omega /c$   rather close to $\mu$,
the lifetime of the constructed cloud is not very long and its energy content is tiny in comparison to
that of the black hole. By choosing $\omega/c$ even closer to $\mu$ we can make the cloud 
more long-lived and more energetic. Again, then the differences between the three types of
black hole spacetimes are even smaller than in our example. So we may conclude that, for all
cases of possible astrophysical relevance, it is not possible to discriminate between the 
three different types of charged black holes with the help of quasi-bound states of uncharged
scalar fields.

\section{The Thomas--Fermi approximation for scalar quasi-bound states}
\label{sec:tf}

Finally, we want to investigate to what extent the Thomas-Fermi approximation
can be used for modeling the quasi-bound scalar field configurations we have
constructed in the preceding section. The Thomas-Fermi  approximation is often 
used for describing the behavior of Bose--Einstein condensates, see for 
instance Ref.\,\cite{pethick2002}. Approximate solutions for scalar field 
distributions in a curved spacetime have been obtained with this method
in Refs.\,\cite{NOS,NOS1} for the  cases of Schwarzschild and Schwarzschild--de Sitter 
background spacetimes. We also refer to Ref.\,\cite{SouzaPires2014} where the validity 
of the Thomas-Fermi approximation was demonstrated for dark matter halos using the 
non-relativistic Gross-Pitaevskii equation. 

The Thomas-Fermi approximation assumes that the kinetic energy 
is negligibly small in comparison to the potential energy and the self-interaction
energy. Then the first term in (\ref{eq:KGGP}) can be neglected and (\ref{eq:KGGP})
can be algebraically solved for $|u(r)|^2$,
\begin{equation}
\label{TF}
|u(r)|^2 = \left( \frac{\omega^{2}}{c^2}-V_{\rm eff} (r) \right) \dfrac{r^2}{\lambda_{\rm eff} (r)} \, .
\end{equation}
Of course, this equation is meaningful only as long as the right-hand side
is positive. Iif $\mu$ has been chosen such that $V_{\mathrm{eff}}$
admits a minimum, this is true on a finite interval $r_1<r<r_2$ around 
the minimum if (\ref{eq:mucon}) holds. 
Outside of this interval one sets $u(r)$ equal to zero. So in the 
Thomas-Fermi approximation the condensate occupies a spherical shell of
inner radius $r_1$ and outer radius $r_2$, where $r_1$ and $r_2$ are the 
solutions of the equation $V_{\mathrm{eff}} (r) = \omega ^2/c^2$. If one plugs the
function 
\begin{equation}\label{eq:TF2}
u(r) = \left\{
\begin{matrix}
\left( \dfrac{\omega^{2}}{c^2}-V_{\rm eff} (r) \right) \dfrac{r^2}{\lambda_{\rm eff} (r)}
\quad \text{if} \quad 
 \dfrac{\omega^{2}}{c^2} > V_{\rm eff} (r) 
\\[0.4cm]
\qquad 0 \qquad \qquad \qquad \quad \text{otherwise}
\end{matrix}
\right.
\end{equation}
into the Gross-Pitaevskii--type equation (\ref{eq:KGGP}), one finds that the
first term diverges if one of the boundary values, $r_1$ or $r_2$, is 
approached. Similarly, the energy density (\ref{eq:Ttt}) diverges at $r_1$ and
at $r_2$. The total energy, on the other hand,  remains finite and the 
Thomas-Fermi approximation has proven very
useful for estimating the size and the total energy of (quasi-)bound states.
For a class of potentials that include the oscillator potential, but \emph{not} our
potential $V_{\mathrm{eff}}$, it has been rigorously proven \cite{LiebSeiringerYngvason2000}
that the Thomas-Fermi approximation becomes arbitrarily good if $\lambda \, N$ 
becomes sufficiently big.

\begin{figure}[!hbt]
\centering
\includegraphics[width=0.75\textwidth,origin=c,angle=0]{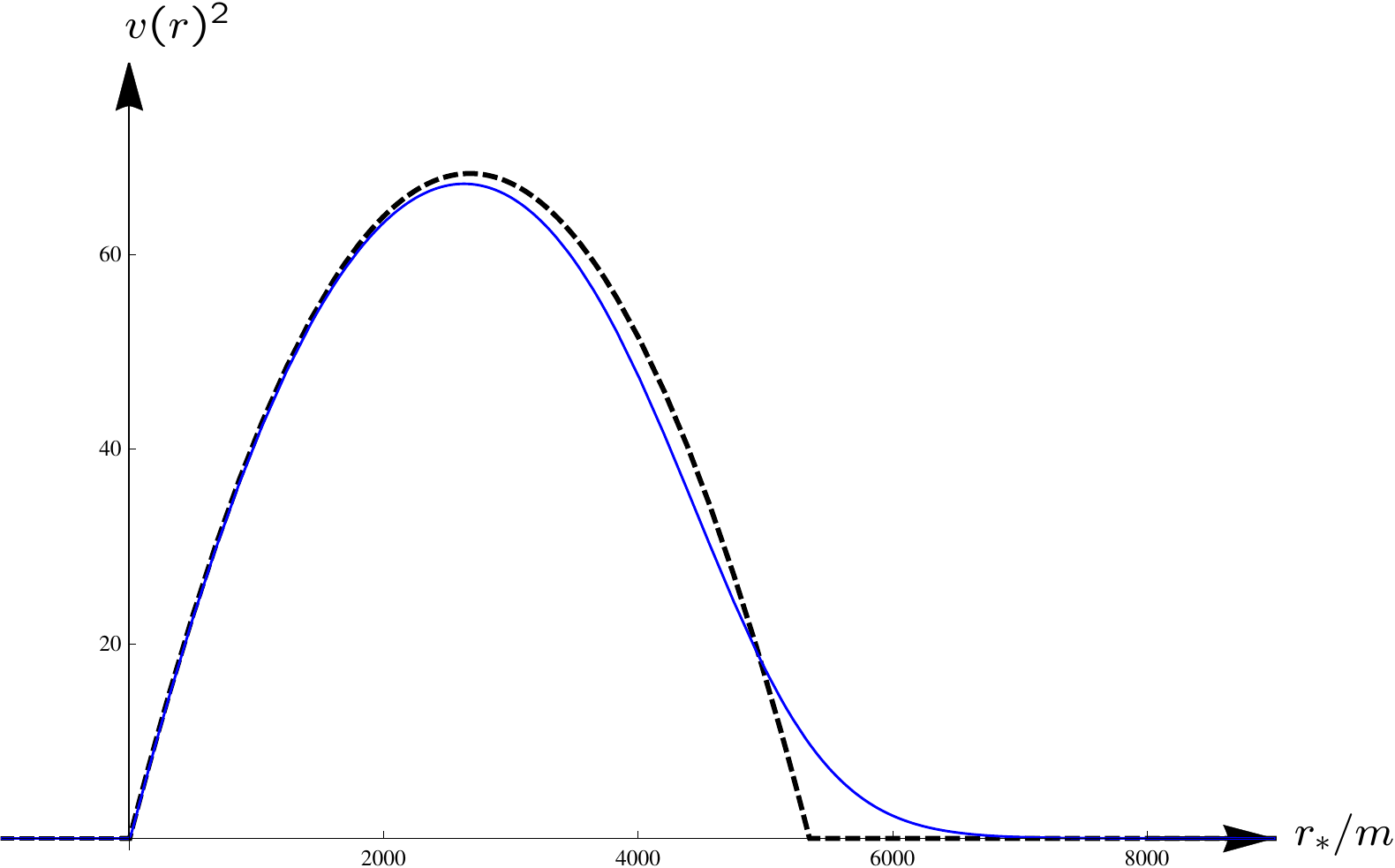}
\caption{\label{fig:TF} Density function $v(r)^2 = \lambda u(r)^2$ 
in a RN black hole, plotted against the tortoise coordinate, for a 
numerical solution of (\protect\ref{eq:KGGP2}) with $q=0.634m$, 
$\mu = 0.16 m^{-1}$ and $\omega /c = 0.15997 m^{-1}$ (solid)
and for the Thomas-Fermi approximation (dashed).
}
\end{figure}

As, in our case, $\lambda \, N$ becomes big if $\omega /c$ is chosen close to
$\mu$, one may expect that the Thomas-Fermi approximation is good if
we choose $\mu$ correspondingly. Our numerical studies have shown that 
one has to choose $\omega /c$ \emph{very} close to $\mu$ for getting a
good agreement between the Thomas-Fermi approximation and the 
exact (numerical) solution. Fig.~\ref{fig:TF} shows the result for the
parameters that have already been used in the preceding section. One
sees that the approximation gives the correct order of magnitude, but
that there is a considerable deviation in the outer part of the cloud. 
The difference becomes
smaller if $\omega /c$ is chosen closer to $\mu$, but the difference
between the Thomas-Fermi approximation and the exact (numerical)
solution is, in any case, bigger than the differences between the
three charged black hole spacetimes, compare Fig.~\ref{fig:TF} with
Fig.~\ref{fig:3wigs}.

\section{Conclusions and outlook}
\label{sec:conclusions}

We have analyzed uncharged scalar test fields that satisfy a Klein-Gordon
equation with a self-interaction term on 
spacetimes of different charged black hole models. We have chosen
the mass parameter $\mu$ of the scalar field such that the effective
potential admits a local minimum which allows for approximate
solutions that depend on time only via a factor $e^{i \omega t}$ with a 
\emph{real} frequency $\omega$. These solutions may be viewed as
quasi-bound clouds of a Bose-Einstein condensate  
around the charged black hole. Mathematically, they come about 
by replacing the effective potential with a cut-off potential that
prevents particles from tunneling through the potential barrier towards
the horizon. The (stationary) solutions in the cut-off potential
may be viewed as good approximations of (non-stationary, decaying)
solutions in the original potential if the tunnel current is sufficiently
small.  It was our main
goal to find out whether or not the density distribution
of such a quasi-bound cloud is different for different charged black-hole
models. We have found that for the three types of black holes
considered here -- the Reissner-Nordstr{\"o}m black hole, Hoffmann's
Born-Infeld black hole and the regular Ay{\'o}n-Beato--Garc{\'\i}a black
hole -- the differences are tiny. 

The type of clouds we have considered here exists only for 
very light bosonic particles. We have seen that, for a supermassive 
black hole with mass parameter $m = GM/c^2 > 10^6 \mathrm{km}$, 
the particle mass $M_{\Phi}$ cannot be bigger than
$10^{-14} \mathrm{eV}/c^2$ because otherwise the effective potential
does not have a minimum. We have also seen that we need a
fine-tuning between the frequency $\omega$ and the mass parameter
$\mu$ of the scalar field if we want to get a cloud with a lifetime
that is long enough for being astrophysically relevant and with 
an energy content that is not completely negligible in comparison
to the energy of the black hole. If the energy content of the 
cloud is comparable to the energy of the black
hole, or even bigger, the cloud may be actually observed,
e.g., with the help of lensing. Of course, for such heavy clouds 
we cannot use the test-field approximation anymore.

We have considered an \emph{uncharged} scalar field because in
this paper we wanted to concentrate on gravitational effects. For a 
charged scalar field the situation is  different, because of 
the electromagnetic interaction between the black hole and the cloud.
It is possible that this electromagnetic interaction gives rise
to observable effects that may be used for discriminating between
different types of charged black holes.


\acknowledgments

This work was supported by DFG-CONACyT Grant Nos.\,B330/418/11 and 211183, by
CONACyT Grant No. 166041F3.
E.C. acknowledges MCTP for financial support. C.L. and V.P. acknowledge support 
from the DFG within the Research Training Group 1620 \emph{Models of Gravity} and 
C.L. also within the QUEST Center of Excellence.


\end{document}